\documentclass[aps,prd,nofootinbib,preprint]{revtex4-1}

\usepackage{amsmath,amssymb,color,graphics,graphicx}
\usepackage{bm}
\usepackage{hyperref} 

\definecolor{dark-green}{rgb}{0,0.7,0}
\definecolor{dark-blue}{rgb}{0,0.2,0.5}
\definecolor{med-blue}{rgb}{0,0.7,1}
\definecolor{mblue}{rgb}{0,0.2,1}
\definecolor{cnc}{rgb}{0.8,0,0}
\definecolor{light-red}{rgb}{1,0.8,0.8}
\definecolor{dark-yellow}{rgb}{1,0.8,0}
\definecolor{light-blue}{rgb}{0.8,0.9,1}
\definecolor{verylight-blue}{rgb}{0.93,0.95,1}
\definecolor{light-yellow}{rgb}{1,0.9,0.8}
\definecolor{grey}{gray}{0.88}

\usepackage{amsfonts}

\begin{document}

\title{Hyperfluid model revisited}

\author{Yuri N. Obukhov}
\affiliation{Russian Academy of Sciences, Nuclear Safety Institute, 
B.Tulskaya 52, 115191 Moscow, Russia
\email{obukhov@ibrae.ac.ru}}

\author{Friedrich W. Hehl}
\affiliation{Institute for Theoretical Physics, University of Cologne, 50923 Cologne, Germany
\email{hehl@thp.uni-koeln.de}}

\begin{abstract}
A new variational theory of a hyperfluid is constructed which is free of the supplementary condition issue and provides a consistent model of the classical matter source in the metric-affine gravity theory. 
\end{abstract}

\maketitle

\section{Introduction}

Among the variety of so-called modified gravitational theories, metric-affine gravity (MAG) is the most natural extension of Einstein's general relativity (GR) theory \cite{Hehl:1976:1,Hehl:1976:2,Hehl:1976:3,Hehl:1977,Ponomarev:1982}. It admits a consistent gauge-theoretic formulation as a gauge theory of the general affine group $GA(4,R)$, which is a semidirect product of the general linear group $GL(4,R)$, and the group of local translations \cite{Hehl:1995}. The gravitational gauge potentials are identified with the metric, the coframe, and the linear connection, whereas the corresponding gravitational field strengths are the nonmetricity, the torsion, and the curvature, respectively. MAG takes into account microstructural properties of matter (such as spin, dilation, and shear currents) as possible physical sources of the gravitational field on an equal footing with macroscopic properties (energy and momentum) of matter. 

Leaving aside the construction of the dynamical scheme for MAG, that encompasses the choice of the Lagrangian of the gravitational field and the analysis of the corresponding field equations, we focus here on the study of the physical sources of  metric-affine gravity. The microscopic matter can be described in terms of manifields (world tensors or world spinors) which realize irreducible representations of the general linear group \cite{Hehl:1995,Neeman:1996}{\color{red};} however, the fundamental wave equation{\color{red},} analogous to the Dirac equation{\color{red},} is still not established. On the other hand, the physical nature of macroscopic matter is much better understood in the framework of the continuous mechanics of media with microstructure \cite{Mindlin:1964,TT,capriz,Eringen99,Lichnerowicz,Volovik:1980,Volovik:2009,Gronwald:1993}. 

The spin fluid (also known in the literature as ``spinning fluid'') represents a special case of continuous matter with microstructure \cite{WR,corben,Ray:1982,Kopc,OK,OP}, the dynamics of which is satisfactorily described in the framework of the Cosserat approach \cite{coss,halb}. The elements of such media are characterized by a rigid local material frame, representing the degrees of freedom of an intrinsic rotation, or spin, of matter elements. The hyperfluid model was developed in \cite{Obukhov:1993,Obukhov:1996} as a natural extension of the concept of an ideal spinning fluid to the case of a deformable material frame, thus adding the intrinsic dilation and shear degrees of freedom. The hyperfluid model attracted considerable attention in the analysis of the dynamics of micromorphic hyperelastic continua \cite{capriz:2002,capriz:2005,capriz:2007}, whereas in the gravity theory it was mostly used in the cosmological context \cite{Smalley:1995,Babourova,Puetzfeld:2001,Obukhov:2015,Iosifidis:2020,Iosifidis:2021}.

A peculiar feature of the original model \cite{Obukhov:1993} was the ``built-in'' generalized Frenkel-Pirani supplementary condition which imposes quite strong restrictions on the structure of the hypermomentum. Thereby it essentially reduces the space of possible gravitational field configurations in the cosmological MAG models. In \cite{Obukhov:1996} an attempt was made to avoid the Frenkel-Pirani type condition. However, the resulting unconstrained hyperfluid model is characterized by a trivial dynamical law for the hypermomentum current that ultimately leads to a decoupling of the hypermomentum from the energy-momentum current.

Here we construct a new variational framework for a hyperfluid model in which the issue of the supplementary condition is resolved. Technically, this is achieved by allowing for an arbitrary four-dimensional intrinsic hypermomentum density without artificially restricting it to the three-dimensional form. 

Our basic notations and conventions follow \cite{Hehl:1995,Birkbook}. In particular, the indices from the middle of the Latin alphabet $i,j,k,\ldots = 0,1,2,3$ label the 4-dimensional spacetime components, the Greek alphabet is used for anholonomic frame indices $\alpha, \beta, \ldots = 0,1,2,3$, whereas the capital Latin indices from the beginning of the alphabet $A,B,C,\ldots = 1,2,3$ refer to the 3-dimensional objects and operations.

\section{Elementary Lagrange-Noether machinery of MAG}\label{MAG}

The geometrical arena of metric-affine gravity theory is the four-dimensional smooth manifold which is endowed with a metric $g_{ij}$, and a linear connection $\Gamma_{ki}{}^j$. In general, these fundamental geometrical objects are completely independent, and the spacetime geometry is exhaustively characterized by the tensors of the curvature, the torsion{\color{red},} and the nonmetricity, respectively:
\begin{align}
R_{kli}{}^j &:= \partial_k\Gamma_{li}{}^j - \partial_l\Gamma_{ki}{}^j + \Gamma_{kn}{}^j \Gamma_{li}{}^n - \Gamma_{ln}{}^j\Gamma_{ki}{}^n,\label{curv}\\
T_{kl}{}^i &:= \Gamma_{kl}{}^i - \Gamma_{lk}{}^i,\label{tors}\\ \label{nonmet}
Q_{kij} &:= -\,\nabla_kg_{ij} = - \partial_kg_{ij} + \Gamma_{ki}{}^lg_{lj} + \Gamma_{kj}{}^lg_{il}.
\end{align}
Physically, one can view ($g_{ij}, \Gamma_{ki}{}^j$) as the gauge gravitational potentials, whereas (\ref{curv})-(\ref{nonmet}) are treated as the corresponding gravitational field strengths \cite{Hehl:1995}.

Denoting arbitrary matter field $\Psi^{\mathcal A}$, where separate matter variables are labeled by the collective index ${\mathcal A}$ without specifying the tensor/spinor nature of these variables, we derive the dynamical field equations for the matter from the action 
\begin{equation}\label{action}
I = \int d^4x\sqrt{-g}\,L^{\rm m}(\Psi^{\mathcal A}, \nabla_i\Psi^{\mathcal A}, g_{ij}).
\end{equation}
This action describes the matter minimally coupled to gravity, hence the connection $\Gamma_{ki}{}^j$ enters only through the covariant derivative $\nabla_i$, whereas the matter Lagrangian $L^{\rm m}$ does not depend on curvature, torsion{\color{red},} and nonmetricity (\ref{curv})-(\ref{nonmet}). 

The canonical energy-momentum tensor and the hypermomentum tensor are defined as
\begin{align}\label{Scan}
\Sigma_k{}^i &= {\frac {\partial L^{\rm m}}{\partial\nabla_i\Psi^{\mathcal A}}}\,\nabla_k\Psi^{\mathcal A}
- \delta_k^i\,L^{\rm m},   \\
\Delta^i{}_j{}^k &= {\frac {\delta L^{\rm m}}{\delta \Gamma_{ki}{}^j}}.\label{Dcan}
\end{align}
In addition, the metrical energy-momentum tensor is determined from the variation of the action with respect to the spacetime metric:
\begin{align}
\sigma^{ij} := -\,{\frac 2{\sqrt{-g}}}{\frac {\delta (\sqrt{-g}L^{\rm m})}{\delta g_{ij}}}.\label{tij0}
\end{align}

The standard Euler-Lagrange machinery \cite{hehl76} yields the conservation laws, such as the hypermomentum and the energy-momentum conservation laws, respectively \cite{Obukhov:2015}:
\begin{align}\label{tskew}
{\stackrel * \nabla}{}_j \Delta^i{}_k{}^j &= \Sigma_k{}^i - \sigma_k{}^i,\\
{\stackrel * \nabla}{}_i\Sigma_k{}^i &= \Sigma_l{}^i T_{ki}{}^l - \Delta^m{}_n{}^l R_{klm}{}^n
- {\frac 12}\sigma^{ij}Q_{kij}.\label{cons2b}
\end{align}
Here the modified covariant derivative is defined as ${\stackrel * \nabla}{}_i := \nabla_i - T_{ki}{}^k - {\frac 12}Q_{ik}{}^k$.

\section{Hyperfluid model: variational approach}\label{weyss}

In general relativity, a simple macroscopic material source of the gravitational field is usually modeled as an ideal fluid, the elements of which are structureless particles (i.e., they do not possess either spin nor other internal degrees of freedom). Such a continuous medium (see, e.g., \cite{bailyn,taub,schutz,schutz2,ray72} for the relevant earlier work, for the general discussion of the relativistic ideal fluids{\color{red},} see \cite{anile1,Rezzolla}) is characterized in the Eulerian approach by the fluid 4-velocity $u^i$, the {\it internal energy} density $\rho = \rho(\nu, s)$, the {\it particle density} $\nu$, the {\it entropy} $s$, and the {\it identity (Lin) coordinate} $X$ \cite{lin}. In addition, one assumes that the number of particles is constant and that the entropy and the identity of the elements are conserved during the motion of the fluid. Due to the conservation of the entropy, only reversible processes are allowed. In other words, the variables of the fluid satisfy
\begin{equation}
{\stackrel {*}\nabla}{}_i(\nu u^i) = 0,\qquad u^i\nabla_i X = 0,\qquad u^i\nabla_i s = 0.\label{l3}
\end{equation}

In MAG, the physical sources of the gravitational field are extended to continuous media with {\it microstructure} \cite{capriz,Eringen99,TT} which are characterized by additional variables describing internal properties of the elements of the fluid. An important example of matter with microstructure is provided by the spin fluid \cite{WR,Ray:1982,corben,Kopc,OK,OP} and in particular by liquid $^3$He in the A-phase, see \cite{Volovik:2009,Vollhardt:1990,Vollhardt:2000}.

\subsection{Lagrangian for the hyperfluid}

Following the Cosserat approach \cite{coss}, we describe matter with microstructure as a continuous medium the elements of which are characterized by the 4-velocity $u^i$ and the material triad $b^i_A$, $A=1,2,3$. However, in MAG the latter is not assumed to be rigid, which means that arbitrary deformations of the triad are allowed during the motion of the fluid. Still, we assume that the standard orthogonality and normalization conditions are imposed on the velocity and the triad legs:
\begin{align}\label{ort}
g_{ij}u^iu^j = c^2,\qquad g_{ij}u^ib^j_A = 0. 
\end{align}
The latter condition means that the vectors of the material triad are spacelike. Taken together, the 4-velocity $u^i$ and the material triad $b^i_A$ comprise the {\it material frame} attached to the elements of the fluid,
\begin{equation}
h^i_\alpha = \left\{u^i, b^i_{\hat{1}}, b^i_{\hat{2}}, b^i_{\hat{3}}\right\}.\label{matframe}
\end{equation}
This frame is not orthonormal, and in particular, the scalar products determine a nontrivial $3\times 3$ material metric,  
\begin{equation}\label{ort2}
g_{ij}h^i_A h^j_B = g_{ij}b^i_A b^j_B = g_{AB}(x),\qquad A,B = 1,2,3.
\end{equation}
The metric $g_{AB}(x)$ characterizes the properties of the ``internal geometry'' of the microstructured material elements at each point $x^i$ of spacetime. The material frame $h^i_\alpha$ is different from an arbitrary spacetime frame $e^i_\alpha$ which is identified with the translational gauge potential of metric-affine gravity. The inverse coframe $h_i^\alpha$ is introduced by
\begin{equation}\label{inv}
h_i^\alpha\,h^i_\beta = \delta^\alpha_\beta,
\end{equation}
and from here, with the help of (\ref{ort}), we find explicitly
\begin{equation}\label{matcof}
h_i^{\hat{0}} = {\frac 1{c^2}}u_i,\qquad h_i^A = \Bigl\{h_i^{\hat{1}}, h_i^{\hat{2}}, h_i^{\hat{3}}\Bigr\}\,.
\end{equation}
Formally, we can write the co-triad as $h_i^A := g_{ij}g^{AB}b^j_B$. Since obviously $u^ih_i^A = 0$, and hence
\begin{equation}
u^ih_i^\alpha = \delta^\alpha_{\hat{0}},\label{uhd}
\end{equation}
we find from (\ref{inv}),
\begin{align}
h^i_A h_i^B = \delta_A^B,\qquad h^i_A h_j^A = \delta_j^i - {\frac 1{c^2}}\,u^iu_j.\label{dij}
\end{align}

The evolution of the material frame is encoded in the generalized acceleration tensor:
\begin{equation}\label{Oab}
\Omega^\alpha{}_\beta := h^\alpha_i u^k\nabla_k h^i_\beta = h^\alpha_i u^k\left(\partial_k h^i_\beta
+ \Gamma_{kj}{}^ih^j_\beta\right).
\end{equation}
This is a natural relativistic extension of the angular velocity, with the same dimension $[\Omega^\alpha{}_\beta] = 1/$s. By construction, the object (\ref{Oab}) is a scalar under general coordinate transformations. Its components encompass the fluid's acceleration
\begin{equation}\label{Om0A}
\Omega^{\hat{0}}{}_A = -\,{\frac 1{c^2}}h^i_A\,u^k\nabla_ku_i,\qquad 
\Omega^A{}_{\hat{0}} = h^A_i\,u^k\nabla_ku^i
\end{equation}
(please note that $u^k\nabla_ku_i \neq g_{ij}u^k\nabla_ku^j$), whereas the  $3\times 3$ matrix
\begin{align}
\Omega^A{}_B = h_i^Au^k\nabla_kh^j_B\label{Om}
\end{align}
describes the rotation and deformation of the material triad, measured by an observer comoving with the fluid.

We formulate the hyperfluid model as a natural generalization of the spin fluid model by introducing a {\it specific hypermomentum density} $\mu^\alpha{}_\beta$ carried by the material elements.  Being an extension of the specific spin density, it has the same dimension $[\mu^\alpha{}_\beta] = [\hbar]$. As compared to \cite{Obukhov:1993}, the crucial novelty is that $\mu^\alpha{}_\beta$ is a four-dimensional object, rather than a three-dimensional one. Together with this new {\it microstructural} variable, the physical properties of matter are then characterized by the internal energy density $\rho(\nu, s, \mu^\alpha{}_\beta)$, the particle number density $\nu$, the entropy $s$, and the identity Lin coordinate $X$. As usual, we assume that thermodynamics is encoded in the generalized Gibbs law
\begin{align}
Tds = d\left({\frac {\rho}{\nu}}\right) + p\,d\left({\frac {1}{\nu}}\right)
- {\frac 12}\omega^\alpha{}_\beta\,d\mu^\beta{}_\alpha,\label{gibbs}
\end{align}
where $T$ is the temperature, $p$ is the pressure, and $\omega^\alpha{}_\beta$ is a conjugate to $\mu^\alpha{}_\beta$.

The dynamics of the hyperfluid is governed by the action integral (\ref{action}), where the Lagrangian reads
\begin{equation}\label{L}
L^{\rm m} = -\,\rho(\nu, s, \mu^\alpha{}_\beta) + L^{\rm kin} + L^{\rm c}. 
\end{equation}
The first term generalizes the usual Lagrangian of an ideal fluid, the second term describes the contribution of the kinetic (hypermomentum-deformation) energy,
\begin{equation}\label{Lkin}
L^{\rm kin} = - \,{\frac 12}\nu\mu^\alpha{}_\beta\Omega^\beta{}_\alpha, 
\end{equation}
(check the dimension $[L^{\rm kin}] = {\rm m}^{-3}\,[\hbar]\,{\rm s}^{-1} =\,$[energy density]), whereas the last term collects all the constraints imposed on the fluid's variables by means of Lagrange multipliers:
\begin{equation}\label{Lc}
L^{\rm c} = -\,\nu u^i\nabla_i\lambda_1 + \lambda_2 u^i\nabla_iX + \lambda_3 u^i\nabla_is + \lambda_0
(g_{ij}u^iu^j - c^2) + \lambda^A g_{ij}u^ib^j_A + \lambda^\alpha{}_\beta(h^i_\alpha h_i^\beta - \delta_\alpha^\beta).
\end{equation}
The last constraint conveniently allows to treat the material frame $h^i_\alpha$ and the inverse coframe $h_i^\alpha$ as independent variables. Therefore, the complete set of the physical variables plus the Lagrange multipliers comprise the matter field of the hyperfluid:
\begin{equation}\label{Psi}
\Psi^{\mathcal A} = \left\{u^i, b^i_A, h_i^\alpha, \nu, s, X, \mu^\alpha{}_\beta,
\lambda_0, \lambda_1, \lambda_2, \lambda_3, \lambda^A, \lambda^\alpha{}_\beta\right\}. 
\end{equation}

\subsection{Euler-Lagrange equations}

Variations with respect to the Lagrange multipliers $\lambda_0, \lambda_1, \lambda_2, \lambda_3, \lambda^A, \lambda^\alpha{}_\beta$ yield (\ref{l3}), (\ref{ort}), and (\ref{inv}): in other words, we recover the orthogonality and normalization constraints, together with the conservation of the entropy, the number and the identity of particles during the fluid's motion.

The total variation of the potential energy is easily found by making use of (\ref{gibbs}):\begin{align}
\delta \left[-\,\rho\right] = -\,\delta\nu\,{\frac {\partial\rho}{\partial\nu}} -\delta s\,
{\frac {\partial\rho}{\partial s}} -\delta\mu^\alpha{}_\beta\,{\frac {\partial\rho}{\partial\mu^\alpha{}_\beta}}
= -\,\delta\nu\left[{\frac {\rho + p}{\nu}}\right] - \delta s\left[\nu T\right]
- \delta\mu^\alpha{}_\beta\left[{\frac 12}\,\nu\,\omega^\beta{}_\alpha\right].\label{drho}
\end{align}
Besides the dependence on material variables $\nu, \mu^\alpha{}_\beta, h^i_\alpha, h^\alpha_i$, the kinetic Lagrangian (\ref{Lkin}) also depends on the gravitational field variables $g_{ij}$ and $\Gamma_{ki}{}^j$. The direct computation (see Appendix \ref{app}) yields the total variation (\ref{dLkin}) for the kinetic Lagrangian, and (\ref{dLc}) for the constraint Lagrangian. 

Combining (\ref{drho}), (\ref{dLkin}), and (\ref{dLc}), one then finds for the variations with respect to the fluid variables $X, s, \nu$, respectively:
\begin{align}
{\frac {\delta L^{\rm m}}{\delta X}} &= -\,{\stackrel {*}\nabla}_i(\lambda_2 u^i) = 0,\label{vX}\\
{\frac {\delta L^{\rm m}}{\delta s}} &= -\,{\stackrel {*}\nabla}_i(\lambda_3 u^i) - \nu T = 0,\label{vs}\\
{\frac {\delta L^{\rm m}}{\delta \nu}} &= -\,u^i\nabla_i\lambda_1 - {\frac {\rho + p}{\nu}} 
- {\frac 12}\mu^\alpha{}_\beta\Omega^\beta{}_\alpha = 0.\label{vrho}
\end{align}
In addition, when varying the action with respect to the fluid velocity $u^i$, we need to take into account that $h^i_{\hat{0}} = u^i$ and $h^i_A = b^i_A$. Then we explicitly find
\begin{align}
{\frac {\delta L^{\rm m}}{\delta u^i}} = \lambda^Ag_{ij}b^j_A + 2\lambda_0u_i + \lambda^{\hat{0}}{}_\alpha h^\alpha_i
- \nu\nabla_i\lambda_1 + \lambda_2\nabla_iX + \lambda_3\nabla_is & \nonumber\\
-\,{\frac 12}\nu\mu^\beta{}_\alpha\,h_k^\alpha\nabla_ih^k_\beta + {\frac 12}\nu\mu^{\hat{0}}{}_\alpha u^k\nabla_k h^\alpha_i
+ {\frac 12}\nu h^\alpha_i u^k\partial_k\mu^{\hat{0}}{}_\alpha &= 0,\label{vu} 
\end{align}
whereas the variation with respect to the fluid triad yields
\begin{equation}\label{vb}
{\frac {\delta L^{\rm m}}{\delta b^i_A}} = {\frac 12}\nu h_i^\beta u^k\partial_k\mu^A{}_\beta
+ {\frac 12}\nu \mu^A{}_\beta u^k\nabla_kh^\beta_i + \lambda^A{}_\beta h^\beta_i + \lambda^Au_i  = 0.
\end{equation}
Finally, variations with respect to the inverse material coframe $h^\alpha_i$ and with respect to the fluid's specific hypermomentum density $\mu^\beta{}_\alpha$ results in, respectively,
\begin{align}\label{vinv}
\lambda^\alpha{}_\beta &= {\frac 12}\,\nu\mu^\gamma{}_\beta\Omega^\alpha{}_\gamma,\\
\omega^\alpha{}_\beta + \Omega^\alpha{}_\beta &= 0.\label{vmu}
\end{align}

Contracting the equations of motion (\ref{vu}) and (\ref{vb}) with $u^i$ and making use of (\ref{l3}), (\ref{ort}), (\ref{uhd}), (\ref{vrho}), and (\ref{vinv}), we find the Lagrange multipliers:
\begin{align}\label{l0} 
2\lambda_0c^2 &= -\,\rho - p - {\frac 12}\,\nu\left(u^k\partial_k\mu^{\hat{0}}{}_{\hat{0}} - \mu^{\hat{0}}{}_\gamma
\Omega^\gamma{}_{\hat{0}} + \mu^\gamma{}_{\hat{0}}\Omega^{\hat{0}}{}_\gamma\right),\\
\lambda^Ac^2 &=  -\,{\frac 12}\,\nu\left(u^k\partial_k\mu^A{}_{\hat{0}} - \mu^A{}_\gamma\Omega^\gamma{}_{\hat{0}}
+ \mu^\gamma{}_{\hat{0}}\Omega^A{}_\gamma\right).\label{lA}
\end{align}

Finally, by contracting (\ref{vb}) with the material triad $b^i_A$, we find the equation of motion of the hypermomentum density:
\begin{align}\label{dmu}
{\frac 12}\,\nu\left(u^k\partial_k\mu^A{}_B - \mu^A{}_\gamma\Omega^\gamma{}_B
+ \mu^\gamma{}_B\Omega^A{}_\gamma\right) = 0.
\end{align}
This is similar to the treatment by Delphenich \cite{Delphenich} of the transport of material frame with the help of the generalized acceleration (deformation) tensor (\ref{Oab}).

\subsection{Canonical Noether currents of hypermomentum and energy-momentum}

By definition, the canonical hypermomentum current (\ref{Dcan}) arises from the variation of the action with respect to the connection $\Gamma_{ki}{}^j$. Since the latter enters only the kinetic part of the Lagrangian, we can read off from (\ref{dLkin}) the expression
\begin{align}
\Delta^i{}_j{}^k & = u^k {\mathcal J}^i{}_j\,,\label{dLdG}\\
{\mathcal J}^i{}_j &:= -\,{\frac 12}\nu\mu^\alpha{}_\beta h^i_\alpha h_j^\beta\,.\label{Jij}
\end{align}
Obviously, the dimension $[{\mathcal J}^i{}_j] = [\hbar]/{\rm m}^3$ is the same as for the spin density. We can invert (\ref{Jij}) and find,
\begin{equation}
-\,{\frac 12}\nu\mu^\alpha{}_\beta = h_i^\alpha h^j_\beta {\mathcal J}^i{}_j\,.\label{muab}
\end{equation}
Then it is straightforward to demonstrate that
\begin{equation}\label{dotJ}
-\,{\frac 12}\,\nu\left(u^k\partial_k\mu^\alpha{}_\beta - \mu^\alpha{}_\gamma\Omega^\gamma{}_\beta
+ \mu^\gamma{}_\beta\Omega^\alpha{}_\gamma\right) = h_i^\alpha h^j_\beta \dot{\mathcal J}^i{}_j,
\end{equation}
where the substantial derivative of an arbitrary quantity $\Phi$ along the fluid's flow is defined in the usual way as $\dot{\Phi} := {\stackrel {*}\nabla}_i(u^i\Phi)$.

With the help of (\ref{dotJ}), the Lagrange multipliers simplify to
\begin{align}\label{l01} 
2\lambda_0c^2 &= -\,\rho - p + {\frac 1{c^2}}\,u_iu^j\dot{\mathcal J}^i{}_j,\\
\lambda^Ac^2 &=  h^A_iu^j\dot{\mathcal J}^i{}_j,\label{lA1}
\end{align}
whereas the equation of motion (\ref{dmu}) is recast into a nice tensor form
\begin{equation}\label{eomJ}
\dot{\mathcal J}^i{}_j - {\frac {1}{c^2}}\,u^iu_k\dot{\mathcal J}^k{}_j - {\frac {1}{c^2}}
\,u_ju^k\dot{\mathcal J}^i{}_k + {\frac {1}{c^4}}\,u^iu_ju^lu_k\dot{\mathcal J}^k{}_l = 0.
\end{equation}
  
Substituting the Lagrange multipliers (\ref{l01}) and (\ref{lA1}) into (\ref{vu}), we can derive an important relation that underlies the computation of the canonical energy-momentum,
\begin{align}\label{vu1} 
-\,\nu\nabla_i\lambda_1 + \lambda_2\nabla_iX + \lambda_3\nabla_is - {\frac 12}\nu\mu^\beta{}_\alpha
h^\alpha_k\nabla_ih^k_\beta = u_i\,{\frac {p + \rho}{c^2}} - {\frac 1{c^2}}(g_{ij}u^k - \delta_i^ku_j)
\dot{\mathcal J}^j{}_k.
\end{align}
Let us come back to the constraint Lagrangian (\ref{Lc}). ``On-shell'' (i.e., when the equations of motion are satisfied) it reduces to 
\begin{equation}\label{Lc1}
L^{\rm c} = -\,\nu u^i\nabla_i\lambda_1 = p + \rho + {\frac 12}\nu\mu^\alpha{}_\beta\Omega^\beta{}_\alpha\,.
\end{equation}
With the help of (\ref{vrho}), we find for the fluid Lagrangian (\ref{L}) ``on-shell''
\begin{equation}
L^{\rm m} = p.\label{Lp}
\end{equation}

Consider the canonical energy-momentum tensor (\ref{Scan}) and use of the crucial relation (\ref{vu1}). Then  the hyperfluid Lagrangian (\ref{L}) yields
\begin{align}\label{Sig}
\Sigma_k{}^i &= u^i{\mathcal P}_k - p\left(\delta_k^i - {\frac {u_ku^i}{c^2}}\right),\\
{\mathcal P}_k &= {\frac {\rho}{c^2}}\,u_k - {\frac 1{c^2}}(g_{kj}u^l
- \delta_k^lu_j)\dot{\mathcal J}^j{}_l\,.\label{Pk}
\end{align}
In contrast, the total variation (\ref{dLkin}) leads to the  metric energy-momentum tensor (\ref{tij0}):
\begin{align}
\sigma_k{}^i = {\frac {u_ku^i}{c^2}}\left(\rho + {\frac {1}{c^2}}\,u^lu_j\dot{\mathcal J}^j{}_l\right)
- p\left(\delta_k^i - {\frac {u_ku^i}{c^2}}\right)
- {\frac 1{c^2}}(g_{kj}u^i + \delta_j^iu_k)u^l\dot{\mathcal J}^j{}_l.\label{tij}
\end{align}
We recognize here a generalization of the Belinfante-Rosenfeld relation \cite{Belinfante1,Belinfante2,Rosenfeld}.

It is worthwhile to note that $u^k{\mathcal P}_k = \rho$. For the {\it dust} case (when $p = 0$), we find, from (\ref{cons2b}), the equation of motion for the 4-momentum of matter: 
\begin{equation}
\dot{\mathcal P}_k = -\,{\mathcal J}^i{}_j u^l R_{kli}{}^j + \Sigma_l{}^i T_{ki}{}^l 
- {\frac 12}\sigma^{ij}Q_{kij}.\label{dotP}
\end{equation}
The equation (\ref{dotP}), together with (\ref{eomJ}), represent a generalization of the Mathisson-Papapetrou system for metric-affine spacetime.

Finally, by making use (\ref{Sig}), we can explicitly demonstrate that (\ref{tskew}) is a consequence of the equations of motion for hypermomentum (\ref{eomJ}).

\section{Conclusion and outlook}\label{conc}

We have constructed a new formulation of the hyperfluid model which is free of any supplementary conditions on the hypermomentum current. Recall that in the original model \cite{Obukhov:1993} the latter was subject to the Frenkel-Pirani type conditions
\begin{equation}\label{Frenkel}
{\mathcal J}^i{}_j u^j = 0,\qquad {\mathcal J}^i{}_j u_i = 0, 
\end{equation}
which strongly constrained the possible gravitational field configurations in MAG. One can see that the violation of (\ref{Frenkel}) is directly related to the fact that the intrinsic hypermomentum $\mu^\alpha{}_\beta$ is the four-dimensional object, and in particular $\mu^{\hat 0}{}_A \neq 0$ and $\mu^A{}_{\hat 0} \neq 0$.

An important consequence of removing the condition (\ref{Frenkel}) is the possibility to consider the hyperfluid model of a purely dilatonic (sometimes called Weyl) type when the hypermomentum density reduces to 
\begin{equation}\label{weyl}
{\mathcal J}^i{}_j = {\mathcal J}\,\delta^i_j. 
\end{equation}
Substituting this into (\ref{eomJ}), we find the equation of motion for the dilation charge density
\begin{equation}\label{eomW}
\dot{\mathcal J} = {\stackrel {*}\nabla}_i({\mathcal J}\,u^i) = 0,
\end{equation}
which resembles the electric charge conservation. As a result, we can verify that the canonical (\ref{Sig}) and the metrical (\ref{tij}) energy-momentum tensors coincide:
\begin{equation}\label{SigW}
\Sigma_k{}^i = \sigma_k{}^i = {\frac {\rho}{c^2}}\,u_ku^i - p\left(\delta_k^i - {\frac {u_ku^i}{c^2}}\right).
\end{equation}
Remarkably, however, one can prove that for a wide class of the physically viable MAG models \cite{Hehl:1976:3,Ponomarev:1982,Tres:1995,Tucker:1998,Tucker:1995,Iosifidis:2020,Iosifidis:2021} the dynamics of the gravitational field with the dilatonic hyperfluid is described by the Einstein equations of standard general relativity theory with an effective matter source
\begin{align}\label{eff}
{\stackrel {\rm eff}\Sigma}{}_k{}^i = {\frac {\rho^{\rm eff}}{c^2}}\,u_ku^i
- p^{\rm eff}\left(\delta_k^i - {\frac {u_ku^i}{c^2}}\right),
\end{align}
where the effective energy density and pressure read
\begin{equation}\label{peff}
\rho^{\rm eff} = \rho - \kappa c^2\zeta{\mathcal J}^2,\qquad 
p^{\rm eff} = p - \kappa c^2\zeta{\mathcal J}^2.
\end{equation}
Here $\kappa = 8\pi G/c^4$ is Einstein's gravitational constant, whereas the dimensionless constant parameter $\zeta$ is determined by the structure of the MAG Lagrangian. The dilation charge density ${\mathcal J}$ affects the dynamics of the spacetime metric in the very peculiar way (\ref{peff}), and this potentially may shed new light into an interesting discussion of the possible role of the dilation charge in the context of the dark matter issue \cite{Tres:1995,Tucker:1998,Tucker:1995}, and in the early cosmology. In particular, on the early stages of universe's evolution the dilation charge may avert the cosmological singularity in a similar way to the spin of matter \cite{Trautman:1973,Minkevich:1980,Poplawski:2012,Magueijo:2013}. This qualitatively new feature will be studied in greater detail elsewhere. 

In addition, it is worthwhile to mention that the new formulation avoids the deficiencies of the unconstrained hyperfluid model \cite{Obukhov:1996} where the hypermomentum current is conserved and it is decoupled from the energy-momentum current. All this suggests that the new hyperfluid model provides a physically more reasonable description of the classical matter source in the framework of metric-affine gravity. 

Looking from a broader perspective, possible applications of the hypermomentum concept and the hyperfluid model range from the early cosmology to the heavy ion physics. As soon as a {\it quark-gluon plasma} comes into existence (for the relation between hypermomentum and hadron physics, see \cite{Hehl:1977}), the hyperfluid could be used as a classical approximation at early stages of the universe evolution when, following the inflation at about 10 $\mu s$ after the big bang, the temperature would be about 10$^{12}$ kelvin, see \cite{Huterer}. The use of a spin fluid (see Beccatini \cite{Becattini:2022zvf} and Biswas et al.\  \cite{Biswas:2022bht}) appears to be too restrictive for a quark-gluon plasma, in our  opinion, since an account on the Regge-trajectories-like hadronic excitations is needed. On the other hand, in an interesting study of Floerchinger et al. \cite{Floerchinger:2021uyo} an attempt was made to clarify the relevance of the hypermomentum current in the relativistic fluid dynamics arising in the framework of the quantum effective action formalism. These issues will be further discussed elsewhere.

\appendix

\section{Explicit variations of the fluid Lagrangian}\label{app}

Direct computation of the total variation of the kinetic Lagrangian (\ref{Lkin}) with respect to the matter field variables (\ref{Psi}) yields
\begin{align}
{\frac 1{\sqrt{-g}}}\delta \left[\sqrt{-g}L^{\rm kin}\right] = \delta\nu\left[-\,{\frac 12}\mu^\alpha{}_\beta
\Omega^\beta{}_\alpha\right] + \delta\mu^\alpha{}_\beta\left[-\,{\frac 12}\nu\Omega^\beta{}_\alpha\right] 
+ \delta u^i\left[-\,{\frac 12}\nu\mu^\alpha{}_\beta h_k^\beta\nabla_ih^k_\alpha\right]\nonumber\\
+ \,\delta h^\alpha_i\left[ -\,{\frac 12}\nu\mu^\beta{}_\alpha u^k\nabla_kh_\beta^i\right]
+ \delta h^i_\alpha\left[ {\frac 12}\nu\mu^\alpha{}_\beta u^k\nabla_kh^\beta_i + {\frac 12}\nu h_i^\beta
u^k\partial_k\mu^\alpha{}_\beta + {\frac 12}\mu^\alpha{}_\beta h_i^\beta{\stackrel {*}\nabla}_k(\nu u^k)\right]
\nonumber\\
+\,\delta g_{ij}\left[- \,{\frac 14}g^{ij}\nu\mu^\alpha{}_\beta\Omega^\beta{}_\alpha\right]
+ \delta\Gamma_{ki}{}^j\left[-\,{\frac 12}\nu\mu^\alpha{}_\beta h^i_\alpha h_j^\beta u^k\right].\label{dLkin}
\end{align}
Similarly, for the total variation of the constraint Lagrangian (\ref{Lc}) we derive 
\begin{align}
{\frac 1{\sqrt{-g}}}\delta \left[\sqrt{-g}L^{\rm c}\right] = \delta\lambda_1\left[{\stackrel {*}\nabla}_i(\nu u^i)
\right] + \delta\lambda_2 \left[u^i\nabla_iX\right] + \delta\lambda_3 \left[u^i\nabla_is\right]\nonumber\\
+ \,\delta\lambda_0\left[g_{ij}u^iu^j - c^2\right] + \delta\lambda^A \left[g_{ij}u^ib^j_A\right] + \delta
\lambda^\alpha{}_\beta\left[h^i_\alpha h_i^\beta - \delta_\alpha^\beta\right]\nonumber\\
+\,\delta\nu \left[-\,u^i\nabla_i\lambda_1\right] + \delta X\left[-\,{\stackrel {*}\nabla}_i(\lambda_2 u^i)\right]
+ \delta s\left[-\,{\stackrel {*}\nabla}_i(\lambda_3 u^i)\right]\nonumber\\ 
+ \,\delta u^i\left[-\,\nu \nabla_i\lambda_1 + \lambda_2\nabla_iX + \lambda_3\nabla_is + 2\lambda_0u_i
+ \lambda^A g_{ij}b^j_A\right]\nonumber\\
+ \,\delta b^i_A\left[ \lambda^A u_i\right] + \delta h_\alpha^i\left[ \lambda^\alpha{}_\beta h_i^\beta\right]
+ \delta h^\alpha_i\left[ \lambda^\beta{}_\alpha h^i_\beta\right]\nonumber\\
+\,\delta g_{ij}\left[{\frac 12}g^{ij}\,L^{\rm c} + \lambda_0\,u^iu^j + \lambda^A\,u^ib^j_A\right].\label{dLc}
\end{align}
A technical comment is in order: The result above is obtained with the help of the geometric identity $\partial_i(\sqrt{-g}Au^i) = \sqrt{-g}\,{\stackrel {*}\nabla}_i(A u^i)$ which holds for any scalar function $A$, \cite{Obukhov:2015}; also, see Schouten \cite{Schouten} for a more detailed discussion of covariant derivatives for tensor densities.

\end{document}